\documentclass[aps,pre,longbibliography,twocolumn,floatfix,superscriptaddress,nofootinbib]{revtex4-2}
\pdfoutput=1

\usepackage{preamble}

\def\supplementfilename{sm}

\myexternaldocument{\supplementfilename}

\pdfximage{\supplementfilename.pdf}
\def\numbersupplementpages{\the\pdflastximagepages}

\newif\ifarXiv
\arXivtrue
\begin{document}

\title{Nature of spin glass order in physical dimensions}

\author{Bharadwaj Vedula}
\affiliation{Department of Physics, Indian Institute of Science Education and Research, Bhopal, Madhya Pradesh 462066, India}
\author{M. A. Moore}
\affiliation{Department of Physics and Astronomy, University of Manchester, Manchester M13 9PL, United Kingdom}
\author{Auditya Sharma}
%\email{auditya@iiserb.ac.in}
\affiliation{Department of Physics, Indian Institute of Science Education and Research, Bhopal, Madhya Pradesh 462066, India}
\date{\today}

\begin{abstract}
  We have studied the diluted Heisenberg spin glass model in a
  3-component random field for the commonly-used one-dimensional
  long-range model where the probability that two spins separated by a
  distance $r$ interact with one another falls as $1/r^{2 \sigma}$,
  for two values of $\sigma$, $0.75$ and $0.85$. No de
  Almeida-Thouless line is expected at these $\sigma$ values. The spin
  glass correlation length $\xi_{\text{SG}}$ varies with the random
  field as expected from the Imry-Ma argument and the droplet scaling
  picture of spin glasses. However, when $\xi_{\text{SG}}$ becomes
  comparable to the system size $L$, there are departures which we
  attribute to the features deriving from the RSB picture of spin
  glasses. For the case $\sigma =0.85$ these features go away for
  system sizes with $L >L^*$, where $L^*$ is large
  ($\approx 4000-8000$ lattice spacings). In the case of
  $\sigma = 0.75$ we have been unable to study large enough systems to
  determine its value of $L^*$. We sketch a renormalization group
  scenario to explain how these features could arise. On this scenario
  finite size effects on the droplet scaling picture in
  low-dimensional spin glasses produce some aspects of Parisi's
  replica symmetry breaking theory of the Sherrington- Kirkpatrick
  model.

\end{abstract}

%\pacs{75.10.Nr, 75.50.Lk}

%\keywords{}

\maketitle

\section{Introduction}
\label{sec:Introduction}
The nature of the ordered state of spin glasses has been controversial
for nearly four decades.  Newman and Stein have suggested that there
are at least four possibilities for the ordered state
\cite{newmanstein:03} which are not disallowed by rigorous arguments:
the broken replica symmetry (RSB) picture of Parisi, \cite{PARISI:79,
  parisi:83, rammal:86, mezard1987spin, parisi:08}, the chaotic pairs
state of Newman and Stein \cite{newmanstein:98}, the TNT picture of
Palassini and Young \cite{palassini:00} and Krzakala and Martin
\cite{krzakala:00} and the droplet scaling picture \cite{mcmillan:84a,
  bray:86,FisherHuse:88}.  These different types of ordered states are
distinguished by the free energy cost of droplet excitations whose size is comparable to the size of the system $L$, and the
fractal dimension $d_s$ of the droplet surface. For the RSB picture
and the TNT picture there are excitations of size $L$ which
contain $L^d$ spins in $d$ dimensions with an associated
free energy cost of $O(1)$. In the chaotic pairs and droplet scaling
picture the free energy cost of such excitations increases as
$L^{\theta^{\prime}}$, with $\theta^{\prime} > 0$. For the
RSB and chaotic pairs pictures, the fractal dimension of the droplet
surface has $d_s = d$, so that the droplet surface is space
filling. In the droplet and TNT scaling pictures $d_s < d$. In yet another picture due to Houdayer and Martin \cite{HoudayerMartin1:2000}, which we shall refer to as the \lq\lq mixed" picture, large droplets of order of the system size $L$ have a fractal surface dimension $d_s=d$ and a free energy cost of $O(1)$ as in the RSB picture. In this paper we argue that droplets of size $\mathcal{L}$ much smaller than the system size $L$  have a surface fractal dimension $d_s< d$ and an energy cost which increases as $\mathcal{L}^{\theta^{\prime}}$, just as in the droplet picture, but  when $\mathcal{L} \sim L < L^*$, their energies are of $O(1)$, and their surfaces have then a fractal dimension $d_s = d$, just as in the picture of Houdayer and Martin. However, we shall provide evidence that for system sizes $L> L^*$, where $L^*$ is a large length scale, there is a change of behavior and the droplets behave as expected in the droplet picture and again have  energies which increase as $L^{\theta^{\prime}}$. Thus when $L < L^*$ the appropriate picture of the spin glass ordered state  picture is that of the mixed picture. We use the name \lq\lq mixed" as this picture combines features from both the droplet and RSB pictures. A recent
numerical study of spin glass excitations whose energies are of $O(1)$
in two and three dimensions can be found in Ref. \cite{nussinov:2024}.

The application of a magnetic field helps to differentiate the four
pictures. In the presence of a field there can still be a phase
transition in the RSB picture. This is the de Almeida-Thouless (AT)
transition \cite{AT:78}. At the AT transition the paramagnetic
high-temperature replica symmetric state becomes a state with broken
replica symmetry. There is a similar transition in the chaotic pairs
picture. However, in the TNT picture and the droplet scaling picture
there is no phase transition in a field. Recently we have carried out
simulations of the AT transitions and found evidence that the AT line
goes away below six dimensions \cite{vedula2024evidence}. Then if so,
below six dimensions the ordered state in the thermodynamic limit must be either that of the TNT
picture or that of droplet scaling.

In this paper we present evidence that in low dimensions, such as $d =
3$, the ordered state is that according to droplet scaling, and that
evidence for the RSB picture arises from finite size effects. The  RSB features have been attributed to TNT effects. However, we shall argue that they arise naturally from finite size effects within the droplet scaling picture itself.  We find
that droplets whose size $\mathcal{L}$ is less than the linear size
$L$ of the system have excitation energies which behave as expected on
the droplet picture, increasing as $\mathcal{L}^{\theta'}$.  Finite
size effects invariably complicate numerical studies but in the case
of spin glasses they produce features which have also confused the
search for the correct order parameter. Experiments are not usually
affected by finite size effects, but for them there is the related
issue of making the size of regions which are fully equilibrated large
enough \cite{Alvarez:10}. The droplet scaling picture predicts
according to Fisher and Huse \cite{FisherHuse:88} that droplets at
site $i$ which contain $i$ and which are of scale $\mathcal{L}$ (which
means that they contain more than $\mathcal{L}^d$ spins but less than
$(2 \mathcal{L})^d$ spins), have a free energy ($F$) distribution
$\rho(F,\mathcal{L})$, which in the limit of large $\mathcal{L}$, has
the scaling form
\begin{equation}
  \rho \left( F,\mathcal{L} \right) = \frac{1}{\mathcal{L}^{\theta^{\prime}}} \,
  \tilde{\rho} \left( \frac{F}{\mathcal{L}^{\theta^{\prime}}} \right).
\label{eq:P(F,L)}
\end{equation}
An exponent $\theta$ is usually defined via the variance of the free
energy when the boundary conditions are changed from
$\uparrow \uparrow$ to $\uparrow \downarrow$ or from periodic to
anti-periodic boundary conditions across a system of linear dimension
$L$;
$(F_{\uparrow \uparrow}-F_{\uparrow \downarrow})^2 \sim L^{2
  \theta}$. In the droplet picture the exponent $\theta^{\prime}$ is
equal to the domain wall exponent $\theta$. In the RSB picture
$\theta =d/6$ but $\theta^{\prime}=0$
\cite{Aspelmeier:03,Aspelmeier:16}. Thus in the RSB picture there are
system-wide droplets whose free energy is only of $O(1)$. These
droplets are space filling, that is the fractal dimension of their
interfaces has $d_s=d$.

Eq. (\ref{eq:P(F,L)}) only applies when the droplet scale
$\mathcal{L}$ is smaller than the linear dimension $L$ of the system
(or the size of its fully equilibrated region). When $\mathcal{L}$ is
comparable to $L$ then finite size effects come into play which
produce some of the effects to be found in the RSB picture. In
Sec. \ref{sec:discussion} we explain using a renormalization group
(RG) argument how these finite size effects which arise when the
system's correlation length $\xi_{\text{SG}}$ is of the order of
system size $L$  produce some of the features found in the
Parisi approach to the Sherrington-Kirkpatrick (SK)
\cite{sherrington:75} model. We also explain that if the system size
$L$ is greater than some large length $L^*$ then these RSB features
should disappear. This reflects what is seen in our Monte Carlo
simulations at $\sigma = 0.85$.  In \cite{moore:21} it was suggested
that $L^*$ becomes infinite as $d \to 6$ because the excitations of
free energy $O(1)$ become the excitations of $O(1)$ of the RSB
picture and this question is discussed again in Sec. \ref{sec:discussion}. When $L < L^*$ our RG argument
suggests that the Parisi order parameter function $q(x)$ as $x\to 0$ should have the size dependence  $T/L^{d/6}$ as expected from RSB when modified for finite size effects, rather than the form $T/L^{\theta^{\prime}}$ of the droplet picture  
\cite{Parisi:1980,Aspelmeier:2008}.  In previous studies it has not been
possible to simulate low dimensional systems which have linear
dimensions $L$ large enough so that the  RSB features can be seen to be
finite size effects, but in this paper we have managed to study
systems which do have $L > L^*$. They support the droplet picture
as being the correct description of spin glasses in three dimensions,
with RSB effects arising from finite
size effects.
 
It is only possible to directly study large droplets for
two-dimensional Ising spin glasses by numerical methods
\cite{Weigel:18, hartmann:03, hartmann:04}.  There exist no fast
algorithms which will make studies in higher dimensions very
convincing. In two dimensions there are departures from
Eq. (\ref{eq:P(F,L)}) due to finite size effects which can be
described by conventional finite size scaling arguments, viz that the
standard deviation of the energy of droplets $\Delta E(\mathcal{L})$,
varies as
\begin{equation}
  \Delta E(\mathcal{L}) \sim \mathcal{L}^{\theta'} \left( 1 +B \mathcal{L}^{-\omega} \right),
\label{eq:scalingcorr}
\end{equation}
where $\omega$ is the correction to scaling exponent associated with
the zero-temperature fixed point and $B$ gives the magnitude of this
correction to scaling. In two dimensions the length scale below which
these finite size modifications of Eq. (\ref{eq:P(F,L)}) are
noticeable seems to be of order roughly 60 lattice spacings
\cite{hartmann:04, hartmann:03}. Thus only for droplets larger than 60
lattice spacings across is the simple $\mathcal{L}^{\theta'}$
visible. For domain walls, in two dimensions, asymptopia sets in at
much smaller sizes, presumably due to having a smaller correction to
scaling coefficient $B$.  In early work, using sizes only up to $L =
12$, Bray and Moore \cite{bray:84} found the stiffness exponent
$\theta$ for the size dependence of domain wall excitations to be $ -
0.294 \pm 0.009$. This was close to more recent results
\cite{Weigel:18} using much better methods, which enabled study of
sizes from $L = 8 \text{ to } 10,000$, with the result that $\theta =
-0.2793 \pm 0.0003$. In the metastate study of Hartmann and Young
\cite{hartmann:2024} in two dimensions they found that the large $L$
behavior was reached by $L= 8$ as it was domain walls and not droplets
which determined the approach to asymptopia in their work. This
suggests that it would be worthwhile to extend their work to three
dimensions. In two dimensions the metastate behavior which they found
was that predicted by the droplet picture.

The finite size effects which we believe explain the origin of RSB effects and their disappearance above a
system size $L^*$ is of different origin to that in
Eq. (\ref{eq:scalingcorr}) (see Sec. \ref{sec:discussion}). They arise
because when $\xi_{\text{SG}}$ is of order $L$, the behavior of the
system is largely determined by the $k=0$ mode of the underlying field theory
$q_{\alpha \beta}$, as in the finite size studies of Br\'{e}zin and
Zinn-Justin \cite{BZ:1985}. It is this which produces a parallel with the SK
model. It has been known for a long time that even
in the one-dimensional Ising spin glass model with short-range
interactions that many of the features usually associated with the
Parisi solution of the SK model can be found when the correlation
length approaches the length of the system \cite{Bray:85}.

The RSB effects are visible in quantities such as the Parisi overlap
function $P(q)$ which involves the $k=0$ mode of $q_{\alpha
  \beta}$. There are, however, situations where the important
excitations are \textit{not} of the size of the system. An example of
these arises in the phenomena of temperature chaos
\cite{Bray:87,FisherHuse:88, Aspelmeier:2002}. If the temperature is
changed from a value $T_1$ to $T_2$, (where both are smaller than the
transition temperture $T_c$ ) then above a length scale
$\mathcal{L}_c$ the spin orientations are completely modified, where
\begin{equation}
  \mathcal{L}_c \left( T_1, T_2 \right) \sim \frac{1}{\left( T_2-T_1 \right)^{1/\zeta}}.
\end{equation}
The exponent $\zeta=d_s/2 -\theta'$ (at least on the droplet
picture). This formula is obtained by equating the free energy cost of
flipping all the spins in a region of size $\mathcal{L}_c$, which
scales as $\mathcal{L}_c^{\theta'}$, to the free energy which can be
gained from variations of the droplet surface entropy, which is of
$O((T_2-T_1) \mathcal{L}_c^{d_s/2})$. Values of $\mathcal{L}_c$ in
simulations of three dimensional Ising spin glasses are usually of
only a few lattice spacings (and hence much smaller than the linear
dimensions $L$ of the system), and hence one would expect good
agreement with droplet model scaling ideas. This is indeed the
case. According to \cite{Palassini:2000,Katzgraber:2001}, for the
Ising spin glass in three dimensions $d_s\approx 2.60 (2)$, while
Boettcher \cite{Boettcher:2024} claims $\theta'\approx 0.24(1)$. The
temperature chaos exponent $\zeta$ has been estimated
\cite{Katzgraber:2007} to be 1.04, which is therefore in good
agreement with droplet model expectation based on the value
$\zeta=d_s/2-\theta'= 1.06$. Note that if the values of $d_s$ and
$\theta'$ predicted by replica symmetry breaking, viz $d_s= d =3$ and
$\theta'=0$, then the value of $\zeta$ would be $1.5$, which is much
larger than its observed value. In the case of the SK model the
choices $d_s=d$ and $\theta'=0$ are consistent with the results on
chaos obtained by Parisi and Rizzo \cite{ParisiRizzo:2010}. Because of
the success of the droplet scaling picture in explaining the chaos
length scale when it is small, we shall assume that the droplets of
free energy of $O(1)$  arise only when the droplets
are of the system size $L$ when $L < L^*$, just as in the proposal of Houdayer
and Martin \cite{HoudayerMartin1:2000}. These droplets on the scale of
the system size $L$ are supposed to be \lq\lq sponge-like" and have
$d_s=d$, just like those of the RSB scenario.

In the next section we describe the one-dimensional Heisenberg spin
glass proxy model and our Monte Carlo simulations of it. In
Sec. \ref{sec:discussion} we outline the renormalization group
scenario which is consistent with the existence of TNT/RSB  behavior as a
finite size effect and its apparent disappearance for system sizes
larger than $L^*$. Finally in Sec.~\ref{sec:summary} we review our
results and discuss what remains to be done.

\section{The model and simulations}
\label{sec:model}

We study the same one-dimensional proxy model which we recently used
to determine the behavior of the AT line as $d \to 6$
\cite{vedula2024evidence}. Its Hamiltonian is that of the classical
Heisenberg spin glass in a $3$-component random field
\begin{equation}
  \mathcal{H}=-\sum\limits_{\langle i,j \rangle}J_{ij} \textbf{S}_i \cdot \textbf{S}_j
  -\sum\limits_{i}\textbf{h}_i\cdot\textbf{S}_i \,,
  \label{eqn:vector_sg_hamiltonian}
\end{equation}
where $\textbf{S}_i$ is a classical spin on the $i^{\text{th}}$
lattice site ($i=1,2,\ldots,N$), and is a unit vector of $m=3$
components. The $N $ lattice sites are arranged around a circle. Each
pair of spins $(i,j)$ are separated by a distance $r_{ij}$ and the
geometric distance between a pair of spins $(i,j)$ is given
by~\cite{PhysRevB.67.134410}
\begin{equation}
  r_{ij}=\frac{N}{\pi}\sin\left(\frac{\pi}{N}\left| i-j \right| \right),
  \label{eqn:distance}
\end{equation}
which is just the length of the chord connecting the $i^{\text{th}}$
and $j^{\text{th}}$ spins. The interactions $J_{ij}$ are independent
random variables such that the probability of having a non-zero
interaction between a pair of spins $(i,j)$ falls with the distance
$r_{ij}$ between the spins as a power law:
\begin{equation}
  p_{ij}=\frac{r_{ij}^{-2\sigma}}{\sum\limits_{j\neq i}r_{ij}^{-2\sigma}}.
  \label{eqn:probability}
\end{equation}
If the spins $i$ and $j$ are linked the magnitude of the interaction
between them is drawn from a Gaussian distribution of zero mean and
whose standard deviation is unity, i.e:
$\left[ J_{ij} \right]_{\text{av}}=0$ and
$ \left[ J_{ij}^2 \right]_{\text{av}} = J^2 = 1$.  To generate the set
of interaction pairs~\cite{PhysRevLett.101.107203,sharma2011phase}
$(i,j)$ with the desired probability we pick a site $i$ randomly and
uniformly and then choose a second site $j$ with probability $p_{ij}$.
If the spins at $i$ and $j$ are already connected we repeat this
process until we find a pair of sites $(i,j)$ which have not been
connected. Once we find such a pair of spins, we connect them with a
bond whose strength $J_{ij}$ is a Gaussian random variable with the
above attributes.  We repeat this process exactly $N_b$ times to
generate $N_b$ pairs of interacting spins.

The mean number of non-zero bonds from a site is chosen to be
$\tilde{z}$ (the co-ordination number). The total number of bonds
among all the spins on the lattice is $N_b=N \tilde{z} /2$. When
$\tilde{z}=6$ this feature is that also of the 3D simple cubic lattice
model and this value for $\tilde{z}$ was used for both the $\sigma$
values studied.  For $\sigma=0$ and $\tilde{z}=N-1$, the model becomes
the infinite-range Sherrington-Kirkpatrick (SK) model
\cite{sherrington:75}, where the zero-field transition temperature is
$T_c =\sqrt{\tilde{z}}/m$. The Cartesian components $h_i^{\mu}$ of the
on-site external field are independent random variables drawn from a
Gaussian distribution of zero mean with each component having variance
$h_r^2$.

Models similar to this have already been studied
\cite{Kawamura:10,sharma2011phase}. Even though it involves spins of
$m$ (=3) components, its AT transition (when it exists) is in the
universality class of the Ising ($m=1$) model
\cite{sharma2010almeida}. Despite the additional degrees of freedom of
the spins compared to those of the Ising model, the Heisenberg model
is easier to simulate than the Ising model as the vector spins provide
a means to go around barriers rather than over them as in the Ising
case, allowing larger systems to be simulated \cite{lee2007large}.  We
ourselves have extensively studied the XY ($m=2$) version of it
\cite{XY:23}, when we concentrated mainly on cases with
$\sigma = 0.75$ and $\sigma=0.85$. Since writing that paper we have
discovered that the Heisenberg case ($m=3$) runs slightly faster, and
so in this paper we have studied these same values of $\sigma$ but for
the Heisenberg model. The results relating to TNT effects have turned
out to be strikingly similar to those found for the XY model
\cite{XY:23}, even though the exponent which describes the behavior of
the correlation length differs (see Table
\ref{tab:scaling_exponents}). The mapping from $\sigma$ to a finite
dimensionality $d_{\text{eff}}$ is complicated when $\sigma > 2/3$,
but, in Ref. \cite{Banos:12}, it was argued that the Ising version of
our system with $\sigma = 0.896$ corresponded to $d_{\text{eff}}=3$,
while $\sigma =0.790$ corresponded to a value of
$d_{\text{eff}}=4$. When $\sigma = 2/3$, $d_{\text{eff}}=6$.

The simulation methods employed in this study follow the approach
outlined in our previous
work~\cite{vedula2024evidence,XY:23,sharma2010almeida}. We use a
combination of overrelaxation and heatbath sweeps, following a 10:1
ratio, where ten overrelaxation sweeps are performed for every
heatbath sweep~\cite{lee2007large,PhysRevB.78.014419}. Overrelaxation
sweeps, also known as microcanonical sweeps, allow the system to
explore microstates at constant energy, while heatbath sweeps enable
proper equilibration by sampling states with different energies
according to the Boltzmann distribution. The parameters for the
simulations, including equilibration times and the number of disorder
samples, are listed in Table~\ref{tab:parameters_fixed_T}.

To verify that the system has reached equilibrium, we apply a test
that leverages the Gaussian nature of the interactions and the
external magnetic field~\cite{PhysRevB.63.184422}. The equilibrium
condition is validated by the relation
\begin{equation}
  \label{eq:equilibration_test}
  U = \frac{\tilde{z}J^2}{2T}\left(q_l-q_s\right)
  + \frac{h_r^2}{T}\left(q - \left|\textbf{S} \right|^2 \right) ,
\end{equation}
where
$U = \frac{1}{N} \left[\langle \mathcal{H} \rangle
\right]_{\text{av}}$ is the average energy per spin,
$q=\frac{1}{N} \sum\limits_i \left[ \left\langle \textbf{S}_i \right
  \rangle \cdot \left\langle \textbf{S}_i \right \rangle
\right]_{\text{av}} $ is the Edwards-Anderson order parameter,
$q_l = \frac{1}{N_b}\sum_{\langle i,j \rangle}
\left[\epsilon_{ij}\left\langle \textbf{S}_i\cdot
    \textbf{S}_j\right\rangle^2\right]_{\text{av}} $ is the ``link
overlap'', and
$q_s = \frac{1}{N_b}\sum_{\langle i,j \rangle}
\left[\epsilon_{ij}\left\langle\left( \textbf{S}_i\cdot
      \textbf{S}_j\right)^2 \right\rangle\right]_{\text{av}}$ is the
``spin overlap'', where $N_b=N\tilde{z}/2$, and $\epsilon_{ij}=1$ if the
$i^{\text{th}}$ and $j^{\text{th}}$ spins are interacting and is zero
otherwise. As the system evolves, the left-hand side (LHS) of this
equation starts small while the right-hand side (RHS) is large, and
the two approach each other from opposite directions as equilibrium is
reached. We calculate both sides of the equation over increasing
numbers of Monte Carlo sweeps (MCSs), with each value doubling the
previous one. The system is considered to be in equilibrium once the
averaged values of Eq.~(\ref{eq:equilibration_test}) match within
error bars for at least two consecutive points. Upon reaching
equilibrium, we transition to the measurement phase to investigate the
system’s properties, with relevant simulation parameters shown in
Table~\ref{tab:parameters_fixed_T}.
 
We focus in this paper on determining and understanding the spin glass
correlation length $\xi_{\text{SG}}$. The wave-vector-dependent
susceptibility is defined via \cite{sharma2011phase}
\begin{equation}
  \label{eq:wvd_sg_susceptibility}
  \chi_{\text{SG}}(k)=\frac{1}{N}\sum\limits_{i,j}\frac{1}{m}\sum\limits_{\mu,\nu}
  \left[\left(\chi_{ij}^{\mu\nu} \right)^2 \right]_{\text{av}}e^{ik(i-j)} ,
\end{equation}
where
\begin{equation}
  \label{eq:chi_ij^mu-nu}
  \chi_{ij}^{\mu\nu}=\left\langle S_i^{\mu} S_j^{\nu} \right\rangle -
  \left\langle S_i^{\mu} \right\rangle\left\langle S_j^{\nu} \right\rangle.
\end{equation}
From it the spin glass correlation length $\xi_{\text{SG}}$ is then
determined using the relation
\begin{equation}
  \label{eq:sg_correlation_length}
  \xi_{\text{SG}}=\frac{1}{2\sin(k_{\text{min}}/2)}
  \left(\frac{\chi_{\text{SG}}(0)}{\chi_{\text{SG}}\left( k_{\text{min}} \right)} -1 \right)^{1/(2\sigma-1)}.
\end{equation}
$k_{\text{min}}= 2 \pi /N$ is the smallest non-zero wavevector
compatible with the boundary conditions. The spin glass susceptibility $\chi_{\text{SG}}=\chi_{\text{SG}}(0)$.
%We have chosen our unit of
%length so that the circumference of the circle, $L$, is equal to $N$.

\begin{figure}  
  \includegraphics[width=0.48\textwidth]{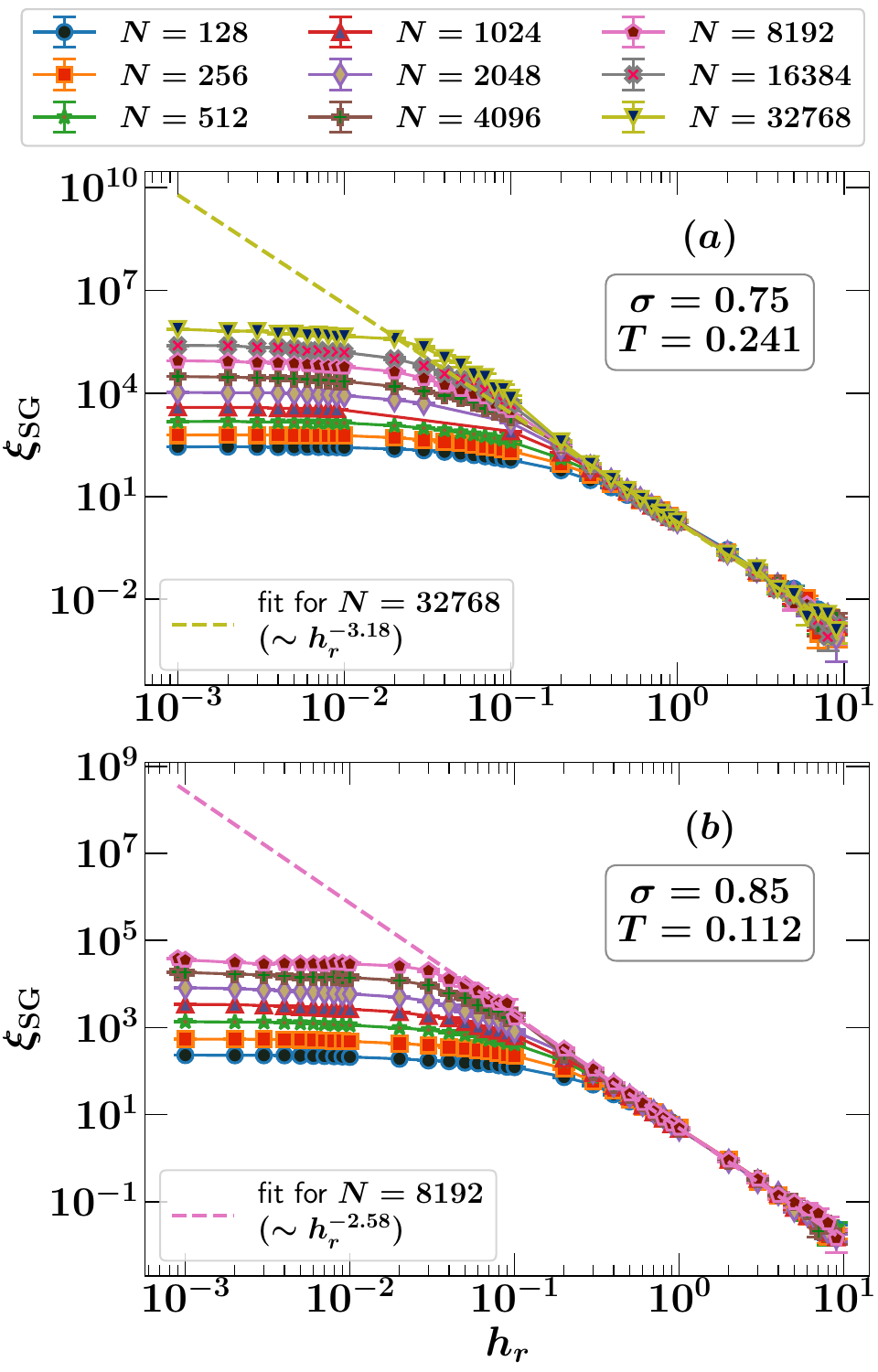}
  \caption{Plots of $\xi_{\text{SG}}$ versus $h_r$ for (a)
    $\sigma = 0.75$ and (b) $\sigma = 0.85$, for a range of system
    sizes $N$. When $\xi_{\text{SG}} \ll N$ the plots become $N$
    independent, as $\xi_{\text{SG}}\sim 1/h_r^x$. For $\sigma =0.75$,
    $x$ is close to 3.18, while for $\sigma = 0.85$, $x$ is close to
    2.58 (see Table~\ref{tab:scaling_exponents}).}
  \label{fig:clfit}
\end{figure}

\begin{figure}  
  \includegraphics[width=0.48\textwidth]{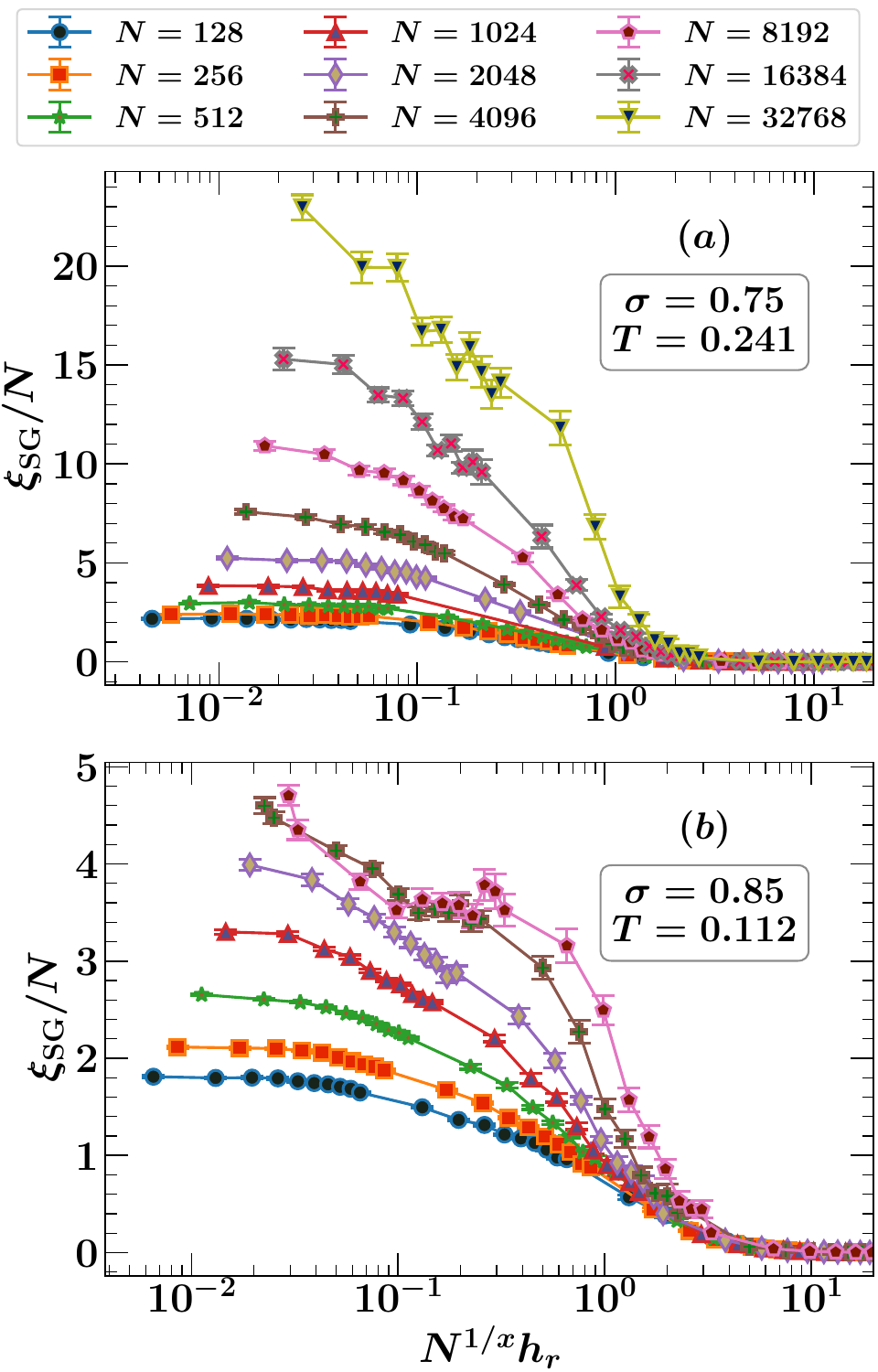}
  \caption{Plots of $\xi_{\text{SG}}/N$ versus $h_r N^{1/x}$ on a
    log-linear scale for (a) $\sigma = 0.75$ and (b) $\sigma =
    0.85$. Note that for the two largest sizes at $\sigma = 0.85$ the
    lines are running close together, suggesting that they are in the
    region where $L > L^*$, but a similar effect is not visible at
    $\sigma = 0.75$, which suggests that in this case $L^*$ is larger
    than the largest system studied at this value of $\sigma$.}
  \label{fig:clNxhr}
\end{figure}

\begin{figure}  
  \includegraphics[width=0.48\textwidth]{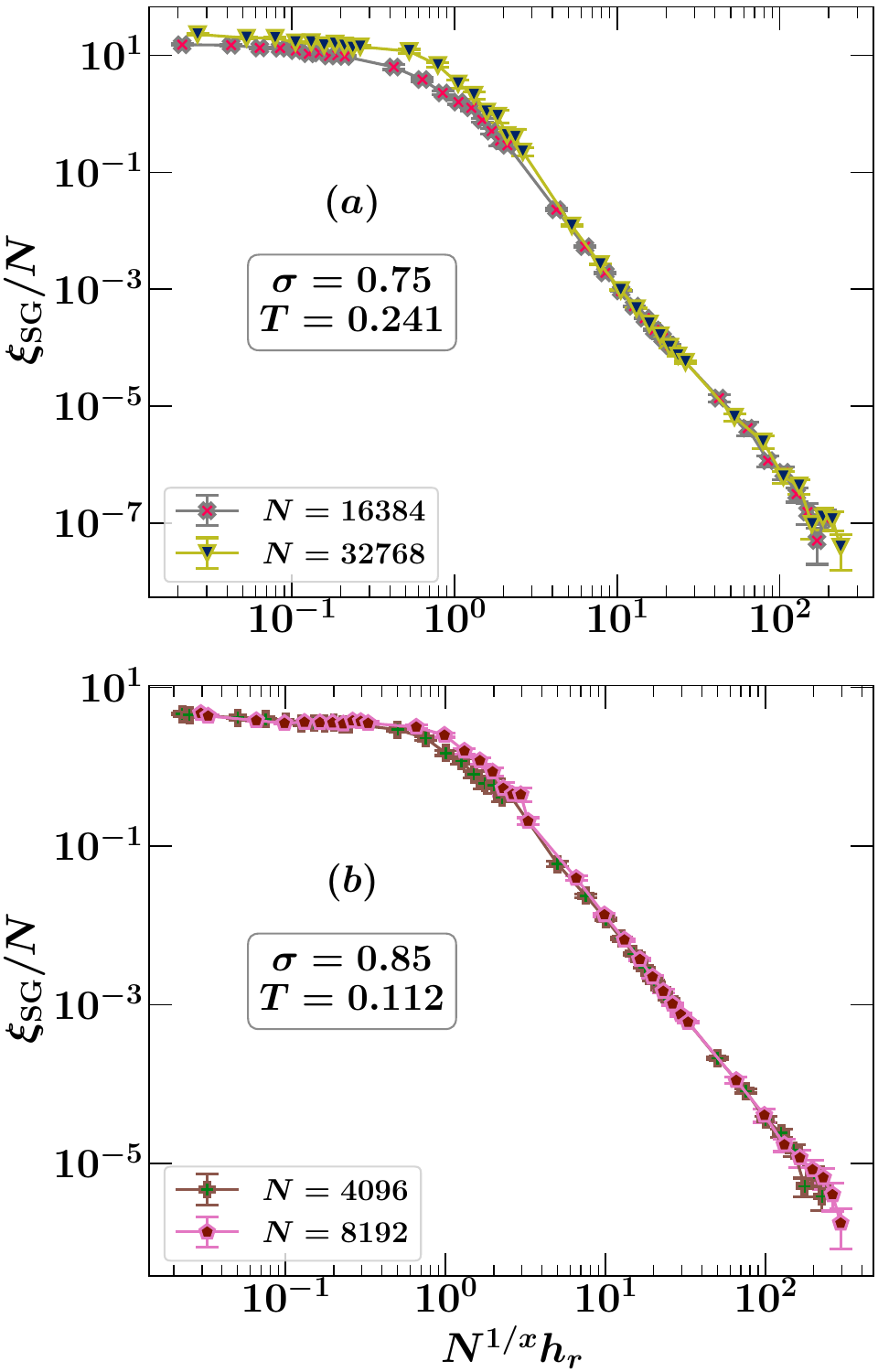}
  \caption{Plots of $\xi_{\text{SG}}/N$ versus $h_r N^{1/x}$ on a
    log-log scale for (a) $\sigma = 0.75$ and (b) $\sigma = 0.85$ for
    the two largest sizes available at each value of $\sigma$. For
    $\sigma = 0.85$ the lines are running close together, suggesting
    that they are in the region where $L > L^*$, and that the data at
    these sizes has the finite size scaling form of
    Eq. (\ref{eq:scalingcorrformxi}). At $\sigma = 0.75$ the curves
    have not run together in the region where $\xi_{\text{SG}}/N$ is
    of $O(1)$, which suggests that in this case $L^*$ is larger than
    the largest system studied and that the finite size form of
    Eq. (\ref{eq:scalingcorrformxi}) is not yet applicable.}
\label{fig:clNxlogN2hr}
\end{figure}

We have studied $\xi_{\text{SG}}$ as a function of $h_r$ at
temperature $T=0.241$ for $\sigma =0.75$, and for $\sigma = 0.85$ at
$T=0.112$, so that for both values of $\sigma$, $T/T_c \approx 0.67$.
$T_c$ is the zero field transition temperature, which for
$\sigma = 0.75$ is $0.359 \pm 0.005$ and for $\sigma = 0.85$ is
$0.166\pm 0.004$, according to \cite{sharma2011phase}. It is possible
to study much larger systems for $\sigma = 0.75$ than for
$\sigma = 0.85$ at $T/T_c=0.67$ as the $T_c$ is smaller at
$\sigma =0.85$ and equilibration of the system is easier at higher
temperatures. We shall now describe the field dependence of
$\xi_{\text{SG}}$ according to the droplet picture
\cite{McMillan_1984, bray:86, FisherHuse:88}, including also the
finite size modifications, and compare these with our simulation
data. We discount the possibility of an AT transition at these values
of $\sigma$ \cite{vedula2024evidence, KY:2005}.

In the droplet picture one uses an Imry-Ma argument
\cite{PhysRevLett.35.1399} for the correlation length $\xi$ and
identifies it with the size of the region or domain within which the
spins become re-oriented in the presence of the random field. The free
energy gained from such a reorientation by the the random field is of
order $\sqrt{q_{\text{EA}}(T)}h_r \xi^{d/2}$. The size of such domains
$\xi$ is determined by equating this free energy to the free energy
cost of the interface of this domain of re-ordered spins with the rest
of the system, which is of the form
$\Upsilon(T) \xi^{\theta^{\prime}}$ ~\cite
{PhysRevE.93.032123}. Equating these two free energies gives
\begin{equation}
  \xi \sim \bigg[\frac{\Upsilon(T)}{\sqrt{q_{\text{EA}}(T)}h_r}\bigg]^{x},
\label{eq:Imry-Malength}
\end{equation}
where 
\begin{equation}
x = \frac{1}{d/2-\theta^{\prime}}.
\label{eqn:xdef}
\end{equation}
While there is a considerable literature on the dependence of the
interface exponent $\theta^{\prime}$ on $\sigma$ for the case of Ising
spin glasses \cite{moore:16}, the case of the Heisenberg spin glass
has hardly been studied. 
Eq.~(\ref{eq:Imry-Malength}) shows that as $h_r\to 0$, the length
scale becomes infinite; $\xi$ diverges as $\xi \sim 1/h_r^x$.
We would expect this formula to apply until finite size effects limit
its growth, which will occur when $\xi$ is of $O(L)$ (or $O(N)$ in our
one-dimensional system). Identifying $\xi_{\text{SG}}$ with $\xi$,
Fig. \ref{fig:clfit} shows that the Imry-Ma fit indeed works well at
the larger fields for both values of $\sigma$; the data for the larger
$h_r$ collapse nicely onto a power law form as predicted by
Eq. (\ref{eq:Imry-Malength}) for all sizes $N$. It only departs from
this formula when $\xi_{\text{SG}}$ becomes of order $N$, when finite
size corrections to the Imry-Ma formula are needed.  RSB effects
produce corrections to the Imry-Ma formula when $\xi_{\text{SG}}$ is
of $O(N)$ unless $N= L > L^*$. But for values of $\xi_{\text{SG}}$ of
the order of just a few lattice spacings, droplet scaling expectations
work very well, just as was also observed for the case of temperature
chaos when the chaos length scale was of the order of a few lattice
spacings \cite{Katzgraber:2007}.

\newcommand{\cgap}{-0.05}
\begin{table*}
  \centering    
  \caption{The table provides the parameters for simulations performed
    at a fixed temperature $T$ with varying field values
    $h_r$. $N(h_r)$ denotes the number of field values sampled within
    the range $h_r$(min,max). Equilibration times differ based on the
    field values and fall within the range
    $N_{\text{sweep}}$(min,max). The number of disorder samples
    collected for various field values is indicated by
    $N_{\text{samp}}$(min,max). $t_{\text{tot}}$ represents the total
    CPU time (in hours) required to generate data for each system
    size. The computation time reflects the total duration needed to
    obtain all the data using a single core with an average clock
    speed of 2.6 GHz. During the measurement phase, one measurement
    was taken every four sweeps.}
  \label{tab:parameters_fixed_T}
  \begin{ruledtabular}
    \begin{tabular}{llrlrrrl}
      % \vspace{1ex}
      $\sigma$  &   $T$ &   $N$  &   $h_r$(min,max) &  $N(h_r)$  &   $N_{\text{sweep}}$(min,max)  &   $N_{\text{samp}}$(min,max) & $t_{\text{tot}}$(hrs) \\[0.15cm]
      \hline
      0.750  &  0.241  &  128  &  $(0.0010,9.0000)$  &  36  &  $(256,2048)$  &  $(4000,16000)$  &  5.03  \\[\cgap cm]
      0.750  &  0.241  &  256  &  $(0.0010,9.0000)$  &  36  &  $(512,4096)$  &  $(4000,16000)$  &  19.67  \\[\cgap cm]
      0.750  &  0.241  &  512  &  $(0.0010,9.0000)$  &  36  &  $(1024,8192)$  &  $(4000,14000)$  &  77.83  \\[\cgap cm]
      0.750  &  0.241  &  1024  &  $(0.0010,9.0000)$  &  27  &  $(2048,16384)$  &  $(1600,14000)$  &  66.30  \\[\cgap cm]
      0.750  &  0.241  &  2048  &  $(0.0010,9.0000)$  &  30  &  $(4096,32768)$  &  $(2000,14000)$  &  391.68  \\[\cgap cm]
      0.750  &  0.241  &  4096  &  $(0.0010,9.0000)$  &  36  &  $(8192,1048576)$  &  $(1600,15000)$  &  8171.79  \\[\cgap cm]
      0.750  &  0.241  &  8192  &  $(0.0010,9.0000)$  &  36  &  $(16384,4194304)$  &  $(663,12800)$  &  40361.13  \\[\cgap cm]
      0.750  &  0.241  &  16384  &  $(0.0010,9.0000)$  &  36  &  $(32768,8388608)$  &  $(240,24725)$  &  129968.71  \\[\cgap cm]
      0.750  &  0.241  &  32768  &  $(0.0010,9.0000)$  &  36  &  $(65536,2097152)$  &  $(240,15280)$  &  326476.90  \\[0.12cm]
      
      0.850  &  0.112  &  128  &  $(0.0010,9.0000)$  &  36  &  $(2048,8192)$  &  $(8000,8000)$  &  18.58  \\[\cgap cm]
      0.850  &  0.112  &  256  &  $(0.0010,9.0000)$  &  36  &  $(4096,524288)$  &  $(4000,18000)$  &  1166.46  \\[\cgap cm]
      0.850  &  0.112  &  512  &  $(0.0010,9.0000)$  &  36  &  $(4096,1048576)$  &  $(2400,20000)$  &  6668.51  \\[\cgap cm]
      0.850  &  0.112  &  1024  &  $(0.0010,9.0000)$  &  36  &  $(16384,4194304)$  &  $(2100,16000)$  &  30521.51  \\[\cgap cm]
      0.850  &  0.112  &  2048  &  $(0.0010,9.0000)$  &  36  &  $(65536,8388608)$  &  $(720,22400)$  &  86440.89  \\[\cgap cm]
      0.850  &  0.112  &  4096  &  $(0.0009,9.0000)$  &  37  &  $(32768,16777216)$  &  $(300,39200)$  &  294691.75  \\[\cgap cm]
      0.850  &  0.112  &  8192  &  $(0.0009,9.0000)$  &  37  &  $(32768,16777216)$  &  $(447,42946)$  &  573696.81  \\
    \end{tabular}
  \end{ruledtabular}   
\end{table*}

The conventional finite size scaling form for corrections to the
scaling of Eq.~(\ref{eq:Imry-Malength}) would be of the form
\begin{equation}
  \frac{\xi_{\text{SG}}}{N} = \mathcal{X} \left( N^{1/x}h_r \right) + N^{-\omega}\mathcal{H} \left( N^{1/x}h_r \right),
\label{eq:scalingcorrformxi}
\end{equation}
where $\omega$ is the correction to scaling exponent, but for the fact
that RSB effects produce large further corrections to these asymptotic
forms, at least when $L < L^*$. As a scaling form
Eq. (\ref{eq:scalingcorrformxi}) should hold at fixed values of $h_r
N^{1/x}$ as $N \to \infty$.  Since in our studies $L^*$ is probably
larger than the length $N$ of our system, at least for $\sigma =
0.75$, the scaling form of Eq.~(\ref{eq:scalingcorrformxi}) does not
work in the region where $\xi_{\text{SG}}$ is of order $N$.  In
Fig. \ref{fig:clNxhr} we have plotted $\xi_{\text{SG}}/N$ against $h_r
N^{1/x}$ on a log-linear plot (which better reveals departures from
scaling) to check the validity of Eq. (\ref{eq:scalingcorrformxi}) for
values of the variable $h_r N^{1/x}$ less than 10. For both values of
$\sigma$ there are clear departures from the scaling form which we
attribute to RSB effects which arise in the region where
$\xi_{\text{SG}} \sim N$. However, for $\sigma =0.85$ where $L^*$ is
expected to be smaller than for $\sigma = 0.75$, Fig. \ref{fig:clNxhr}
hints that Eq.~(\ref{eq:scalingcorrformxi}) might apply as the plots
at adjacent sizes for the larger $N$ values seem to be getting closer
together as $N$ is increased (which is a feature predicted by
Eq.~(\ref{eq:scalingcorrformxi})). The contrasting behaviors for
$\sigma =0.75$ and $\sigma = 0.85$ for the two largest system sizes at
each value of $\sigma$ is displayed in Fig. \ref{fig:clNxlogN2hr} on a
log-log plot so that we can display behavior for all the values of
$h_r N^{1/x}$ which we have studied. We see in
Fig. \ref{fig:clNxlogN2hr} that at $h_r N^{1/x}$ around unity,
Eq. (\ref{eq:scalingcorrformxi}) does seem to apply as $N$ becomes
larger than $4000$ for the case of $\sigma = 0.85$. However, for the
case when $\sigma = 0.75$, where we expect $L^*$ to be larger than for
$\sigma = 0.85$, we do not seem to have reached system sizes $N$ where
droplet scaling will apply.

One complication that we have not discussed is that the scaling form
of Eq. (\ref{eq:scalingcorrformxi}) will not apply for very small
values of the field $h_r$. It is implicit in this equation that the
field $h_r$ is sufficiently large so that, for example, the Parisi
overlap function $P(q)$ is zero for $q < 0$. This requires that
$h_r \sqrt{q_{\text{EA}} N} \gg T$. Plots of $P(q)$ in fields $h_r$ of
$O(1/\sqrt{N})$ can be found in Ref.  \cite{Aguilar-Janita:2024}.  The
argument of the crossover function $h_r N^{1/x}$ at the point of just
satisfying the inequality is of $O \left( N^{(1/x-1/2)} \right)$ which
is small at large $N$ since $x > 2$. In other words, the scaling form
of Eq. (\ref{eq:scalingcorrformxi}) should hold on the droplet picture
for any finite value of $h_r N^{1/x}$ provided $N$ is large
enough. Any departures from Eq. (\ref{eq:scalingcorrformxi}) at the
largest sizes we have studied seem unlikely to be caused by the small
field effect.  The failure to obey droplet scaling forms at
$\sigma = 0.75$ seems more likely to be connected with RSB effects
than with using too small values of $h_r$, as $x$ is larger for this
value of $\sigma$ than for $\sigma = 0.85$, so it should be
\textit{easier} to satisfy the criterion on $h_r$ for this case. Hence
it seems likely the RSB effects are the cause of the poor scaling
collapse in Fig. \ref{fig:clNxhr} for $\sigma = 0.75$, rather than
being caused by using too small values of $h_r$. 

We have also studied the behavior of $\chi_{\text{SG}}$ as a function
of $N$, $h_r$ and $\sigma$ but have relegated its analysis to the
supplementary material because it involves an additional fitting
parameter, $z$, making it less straightforward than the case of
$\xi_{\text{SG}}$. Its behavior though is very similar to that found
for $\xi_{\text{SG}}$.
  
\renewcommand{\cgap}{-0.05}
\begin{table*}[t]
  \centering  
  \caption{Values of the exponents $x_{\chi}$, $x$, and $z$ derived
    from our simulations for various values of $\sigma$ and
    temperature $T$. For Heisenberg spins the values of these
    exponents are obtained from the fitting procedure shown in
    Figs.~\ref{fig:clfit} and \ref{fig:chi-fit}. The corresponding
    values of the exponents for XY spins are obtained from Fig. 9 in
    \cite{XY:23}.}
  \label{tab:scaling_exponents}
  \begin{ruledtabular}
    \begin{tabular}{lccccccc}
      & $\sigma$ & $T$ &  $T_c$ &  $T/T_c$ & $x_{\chi}$ & $x$  & $z$ \Tstrut\\[0.15cm]
      \hline    
      XY  & $0.75$  & $0.55$  &  $0.62$  &  $0.89$  & $1.6114 \pm 0.0005$ & $2.7077 \pm 0.0531$ & $0.5951 \pm 0.0119$ \\[\cgap cm]
      XY  & $0.85$  & $0.3$  &  $0.33$  &  $0.91$  & $1.8919 \pm 0.0014$ & $2.2019 \pm 0.0152$ & $0.8592 \pm 0.0066$ \\[0.15cm]

      Heisenberg  & $0.75$  & $0.241$  &  $0.359$  &  $0.671$  & $1.8926 \pm 0.0003$ & $3.1790 \pm 0.0730$ & $0.5954 \pm 0.0138$ \\[\cgap cm]
      Heisenberg  & $0.85$  & $0.112$  &  $0.166$  &  $0.674$   & $2.1322 \pm 0.0011$ & $2.5822 \pm 0.0387$ & $0.8257 \pm 0.0128$ \\
    \end{tabular}
  \end{ruledtabular}
\end{table*}

\section{Origin of RSB Behavior as a Finite Size Effect}
\label{sec:discussion}

We shall now explain using conventional renormalization group (RG)
arguments the origin of RSB behavior as a finite size effect. Our system has a
correlation length $\xi_{\text{SG}}$ which becomes large at small
values of $h_r$. When there is a long correlation length there is
scope for an RG approach. Our simulations show that when the field is
sufficiently small so $\xi_{\text{SG}}$ is of order $L$ then RSB
effects are visible and only get small again if $L > L^*(T, \sigma)$,
where $L^*$ is a large length.  Our results in Figs. \ref{fig:clNxhr}
and \ref{fig:clNxlogN2hr} suggest that $L^*$ for $\sigma= 0.85$ is in
the range $L^* \sim 4000-8000$, while for $\sigma = 0.75$ all we can
say is that is that its $L^*$ is greater than the largest system we
could study, which was $N =32768$.

As $h_r\to 0$, $\xi_{\text{SG}}$ will approach the zero-field cumulant
correlation length, which is of order $L$. This is the origin of the
flattening off of $\xi_{\text{SG}}/N$ in Fig. \ref{fig:clNxlogN2hr} as
$N^{1/x} h_r$ goes to zero. When $T < T_c$, the system at $h_r=0$ is
in the low-temperature phase of the zero-field spin glass. The RG
flows will take the system to its zero-temperature fixed point. In
fact the exponent $\theta'$ which determines $x$ is an exponent of the
zero temperature fixed point in zero-field; it is not associated with
the critical fixed point. In the presence of a non-zero value of
$h_r$, we expect the RG flows to be towards this fixed point, and
veering away from it before it is reached: $h_r$ is a relevant
perturbation at the zero-temperature fixed point.

Under the RG flow, the coupling constants of its field theory will be
flowing towards the values which they take at this (stable)
zero-temperature fixed point (at $h_r=0$). There are, of course, a
large number of such coupling constants but we will just focus on one
of them, the renormalized temperature $T(l)$. Once $\xi_{\text{SG}}$
becomes comparable to its zero field cumulant value then we shall use
the strategy used by Br\'{e}zin and Zinn-Justin \cite{BZ:1985} of
reducing the field theory to just the $q_{\alpha\beta}(k = 0)$
mode. (We are envisaging the use of periodic boundary conditions). The
gradient terms now disappear from the field theory and the remaining
integrals over the $q_{\alpha\beta}$ are then of the same form as is
encountered in the mean-field Sherrington-Kirkpatrick (SK) model
\cite{sherrington:75}. Hence provided one only looks at features which
involve this mode one will see all the features normally seen in that
model such as many states with an ultrametric topology, excitations
which are of $O(1)$ and so on. The neglected modes of non-zero
wavevector make a sub-dominant contribution to the finite size scaling
functions. The Parisi overlap function $P(q)$, which is a function of
$q_{\alpha\beta}(k = 0)$, naturally will have features similar to
those seen in the SK model. It is natural therefore that there will
exist excitations on the scale of the system size $L$ whose free
energies are of $O(1)$ with fractal dimension $d_s=d$, just as was
envisaged by Houdayer and Martin \cite{HoudayerMartin1:2000}, who did
not make use of any RG argument in making their suggestions.

The value to be used for $T(l)$ is its value at which the RG length
scale, which grows as $e^l$, reaches the system size $L$. At this
point mean-field theory will be useful provided $L$ itself is large
enough so that the modes $q_{\alpha \beta}(k)$ with $k$ non-zero can
be ignored.  It is only for an infinite system in zero field that the
coupling constants reach their fixed point values and $T(l)$ becomes
zero. Thus it would seem natural that many of the features associated
with the replica symmetry breaking results of Parisi should appear to
be present in low-dimensional spin glasses, even though they are
essentially present because of these finite size effects. A pronounced
change in behavior once the correlation length becomes of order of the
size of the system was found in the numerical work of
\cite{Manssen:2015}.

For very large systems, the effective temperature $T(l)$ will get very
close to zero. Then the argument of Crisanti and de Dominicis
\cite{Crisanti:2010, Crisanti:2011} will come into play. They observed
that in the SK model as $T\to 0$ the Parisi overlap function $q(x)$
becomes nearly independent of $x$ (until $x \sim T$ when it starts to
fall almost linearly to zero as $x\to 0$). In other words, it is
becoming nearly replica symmetric at low-temperatures.  $q(x)$ is
expected on the droplet picture to be independent of $x$. It has
replica symmetry, i.e. constant $q(x)$. For $L > L^*$ one has reached
effective temperatures $T(l)$ so small that one is in the region
studied in Refs. \cite{Crisanti:2010, Crisanti:2011}. In this limit,
the zero-field Parisi overlap function $P(q)$ at $q=0$ will decrease
as $T/L^{\theta'}$, the form expected in the droplet picture
\cite{FisherHuse:88, drossel:98}.  $L^*$ is not associated with any
sharp features but marks the system size above which droplet features
dominate the behavior.

Of course it would be useful if we could demonstrate explicitly these
RG features by direct calculation. For the much simpler problem of the
$\phi^4$ scalar field theory such a program of using RG calculations
to explore finite size features was done in
Ref. \cite{Rulquin:2016}. Alas for spin glasses we do not have
well-founded RG equations for the low-temperature phase although we
have RG equations for the critical behavior in zero field
\cite{harrislubchen:76,Chen:77}. There are real space RG equations
like the Migdal-Kadanoff approximation but they are not consistent
with the existence of many pure states \cite{Angelini:2017} and do not
produce TNT features, as was found long ago \cite{drossel:98} for
temperatures $T$ well below $T_c$. For the Migdal-Kadanoff RG
procedure there is no sign of a large length scale like $L^*$; one quickly
reaches asymptopia. It would therefore be useful if RG equations could
be found which would reach asymptopia only when $L > L^*$, with $L^*$
being a large length scale. They would have to have features which could anticipate the existence of many pure states in order to obtain a large value for $L^*$. While no such RG calculation is known to us,
Middleton~\cite{middleton:13} (see also Ref. \cite{Hatano:2002})
predicted that it would only be at very large values of $L$ that the
decay of the Parisi overlap function $P(q)$ at $q=0$ would be as
$T/L^{\theta'}$ (the form predicted by droplet scaling).

We  mentioned in Sec. \ref{sec:Introduction} that we would expect $L^*$ to approach infinity as $d \to 6$. An argument for this follows from our scenario of RSB behavior for $L < L^*$ through the finite-size dependence of the Parisi order parameter function $q(x)$. In the Parisi approach to RSB, \cite{Parisi:1980}, at the $K$ th level of replica symmetry breaking $q(0)= T/K$, where $c$ is a constant and only vanishes as $K \to \infty$. In Aspelmeier et al. \cite{Aspelmeier:2008} it was argued that in the SK model containing $N$ spins, one only needs to have $K \sim N^{1/6}$  to achieve a stable theory, and this result was used to explain many finite size effects of the SK model. Hence we would expect on the RSB picture that  $q(0) =c^{\prime} T/N^{1/6}$. On the droplet picture $q(0) = c T/L^{\theta^{\prime}}$ and hence droplet behavior will dominate $q(0)$  if 
\begin{equation}
c T/L^{\theta^{\prime}} > c^{\prime} T/L^{d/6}.
\label{eq:L*}
\end{equation}
This implies that droplet scaling behavior will dominate $q(0)$ when $L > L^*$ where $L^*= (c/c^{\prime})^{1/(d/6-\theta^{\prime})}$.  As the exponent $1/(d/6-\theta^{\prime}) \to \infty$ as $d \to 6$ (based upon Ref. \cite{Boettcher:2024} and the argument that $\theta^{\prime}\to 1$ as $d \to 6$ \cite{vedula2024evidence}),
then $L^*$ becomes a very large length provided $c > c^{\prime}$. Notice that for $L < L^*$, $q(0)$ on the RSB picture is larger than that predicted by droplet scaling.

\section{Summary and conclusions}
\label{sec:summary}

In this paper, we addressed the long-standing debate regarding the
nature of the ordered state in spin glasses, focusing on four major
theoretical frameworks: the replica symmetry breaking (RSB) picture,
the chaotic pairs state, the TNT picture, and the droplet scaling
picture. Using a detailed simulation study of a one-dimensional spin
glass proxy model, we aimed to clarify the nature of the ordered
state. Specifically, we investigated the droplet scaling and TNT
pictures, where finite size effects often obscure the true behavior of
the system.
%
% Our results suggest that TNT features are a manifestation of finite size effects rather than a
% property of the spin glass phase in the thermodynamic limit.
%
By studying large system sizes, we observed that RSB-like features,
fade away when the system exceeds a critical length scale $L^*$.
These findings suggest that the RSB-like effects are artifacts of
finite size rather than being indicative of the true nature of the
ordered state.  Our results thus provide evidence in favor of the
droplet scaling picture as the correct description of the ordered
state in low-dimensional systems, such as $d = 3$, while highlighting
the role of finite size effects in spin glass simulations.

In our simulations, we utilized a one-dimensional proxy model to
investigate two values of $\sigma > 2/3$: $\sigma = 0.75$ and
$\sigma = 0.85$. The temperature was held constant at approximately
$0.67 \, T_c$, while the magnetic field was varied to generate data
for each $\sigma$ value. For $\sigma = 0.75$, the largest system size
considered was $N=32768$, and for $\sigma = 0.85$, we studied systems
up to $N=8192$, surpassing the maximum system size ($N=4096$) analyzed
in \cite{XY:23} for $\sigma=0.85$. Our findings, illustrated in
Figs.~\ref{fig:clNxhr} and \ref{fig:clNxlogN2hr}, exhibit a notable
data collapse for $\sigma = 0.85$, attributed to the fact that the
largest system size $L$ exceeds the critical value $L^*$ for this
$\sigma$. However, for $\sigma = 0.75$, this collapse was not
observed, as the system sizes explored in this study remain within the
regime where $L < L^*$.

Our numerical results showed that RSB effects appear when
$\xi_{\text{SG}} \sim L$, provided $L < L^*$. These effects vanish
when $L > L^*$. A remaining challenge therefore is to develop detailed
RG calculations that explain the emergence of the $L^*$ feature.
These would hopefully at the same time explain why there appears to be
no AT line below six dimensions \cite{vedula2024evidence} and describe
the dimensionality and temperature dependence of $L^*$. Alas, RG
calculations in the ordered phase of spin glasses are very
challenging. In particular we know of no RG calculation which gives a
large $L^*$. However, the argument given for $L^*$ from Eq. (\ref{eq:L*}) indicates that a large value should come as no surprise.

The scenario outlined in this paper gives an important role to RSB in three dimensional spin glasses when $L < L^*$. Because
$L^*$ is such a large length it seems likely that no simulation has
ever been done in three dimensions with $L > L^*$. Alas the actual
value of $L^*$ for the three dimensional Ising spin glass remains
unknown.

\section*{Acknowledgments}
We thank O. C. Martin, D. L. Stein and A. P. Young for valuable exchanges.
We are grateful to the High Performance Computing (HPC) facility at
IISER Bhopal, where large-scale calculations in this project were
run. B.V is grateful to the Council of Scientific and Industrial
Research (CSIR), India, for his PhD fellowship. A.S acknowledges
financial support from SERB via the grant (File Number:
CRG/2019/003447), and from DST via the DST-INSPIRE Faculty Award
[DST/INSPIRE/04/2014/002461].

% ========== references ====================
\twocolumngrid
\bibliography{refs}
% \input{paper_rev.bbl}g
%==========================================

%========== supplementary material ========================================
\ifarXiv
  

\section*{Supplementary material}

# Supplementary Material:
# Nature of spin glass order in physical dimensions

Bharadwaj Vedula,[1] M. A. Moore,[2] and Auditya Sharma[1]

[1]Department of Physics, Indian Institute of Science Education
and Research, Bhopal, Madhya Pradesh 462066, India

[2]Department of Physics and Astronomy, University of Manchester, Manchester M13 9PL, United Kingdom

## I. SPIN GLASS SUSCEPTIBILITY

The spin glass susceptibility is predicted by the droplet picture to be of the form [44]

$$\frac{\chi_{\text{SG}}}{N^z} = \mathcal{C}(N^{1/x} h_r) + N^{-\omega} \mathcal{G}(N^{1/x} h_r), \tag{S1}$$

where the crossover function $\mathcal{C}(y) \sim 1/y^{x_\chi}$ for large $y$, leading to $\chi_{\text{SG}} \sim \xi^z$ and becoming independent of $N$. In this regime, the susceptibility scales as

$$\chi_{\text{SG}} \sim 1/h_r^{x_\chi}, \tag{S2}$$

implying the relationship $x_\chi = xz$. For small $y$, $\mathcal{C}(y)$ approaches a constant value. Our results, shown in Figs. S1(a) and S1(b), confirm that $\chi_{\text{SG}}$ collapses onto a power law as predicted by this

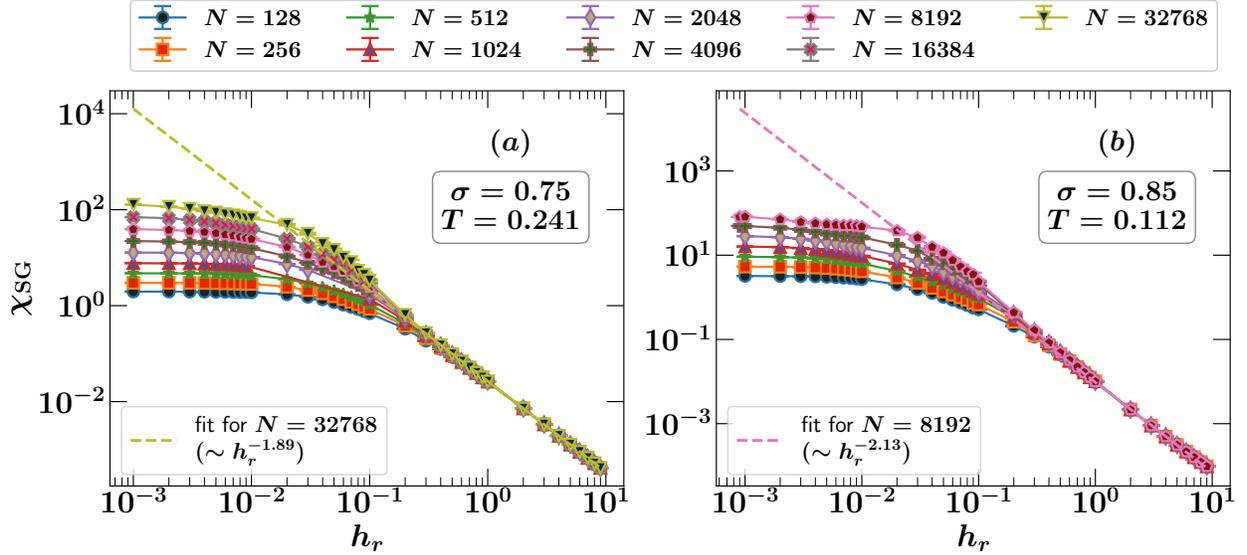

FIG. S1. Log-log plots of $\chi_{\text{SG}}$ as a function of $h_r$ for (a) $\sigma = 0.75$ and (b) $\sigma = 0.85$, shown for various system sizes $N$. At higher values of $h_r$, the curves converge, indicating that $\chi_{\text{SG}}$ becomes independent of $N$, following the scaling behavior $\chi_{\text{SG}} \sim 1/h_r^{x_\chi}$. For $\sigma = 0.75$, $x_\chi$ is approximately 3.18, while for $\sigma = 0.85$, $x_\chi$ is around 2.58 (see Table II).

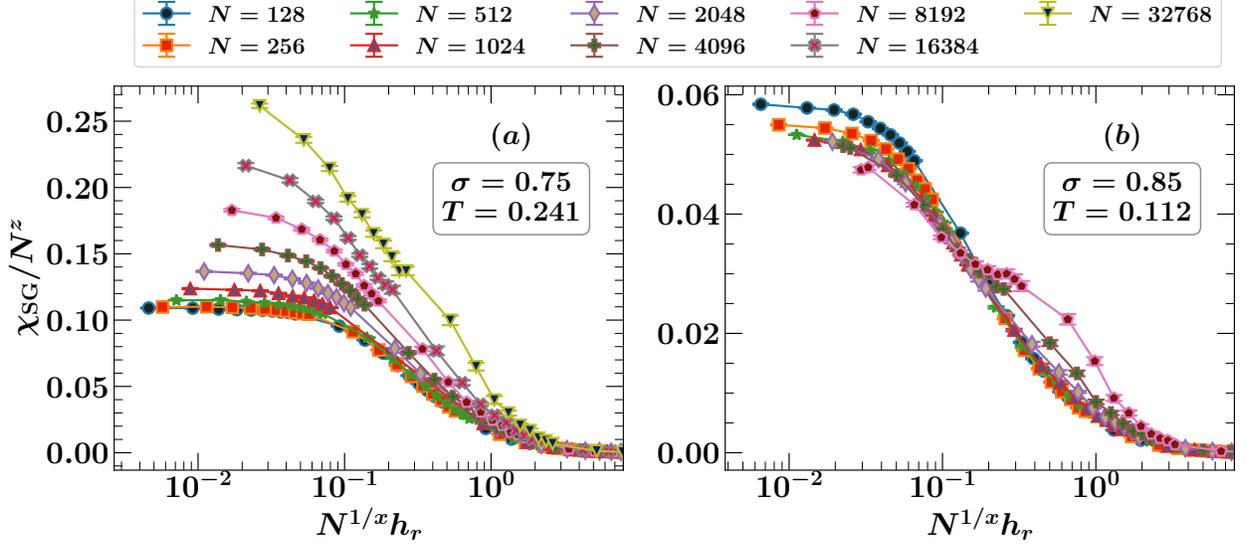

FIG. S2. Plots of $\chi_{\rm SG}/N^z$ against $h_r N^{1/x}$ on a log-linear scale for (a) $\sigma = 0.75$ and (b) $\sigma = 0.85$, illustrating the scaling behavior across different system sizes.

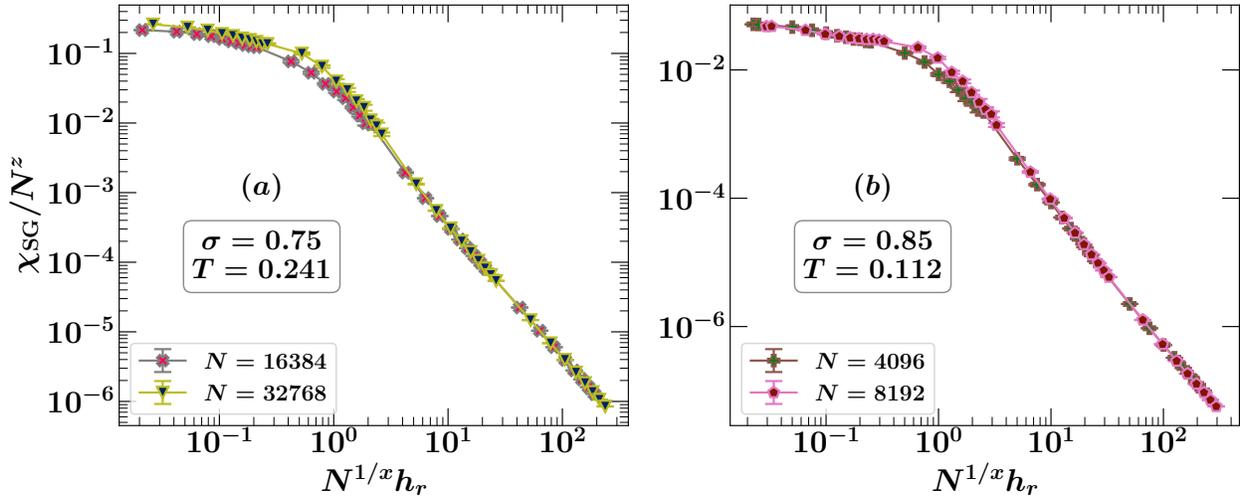

FIG. S3. Plots of $\chi_{\rm SG}/N^z$ versus $h_r N^{1/x}$ on a log-log scale for (a) $\sigma = 0.75$ and (b) $\sigma = 0.85$, showcasing the behavior for the two largest system sizes available at each $\sigma$ value.

form, for all system sizes $N$ when $h_r$ is large. The exponent $z$ must be determined by fitting the data (see Ref. [44] for a detailed discussion of these exponents). Our results for the droplet exponents $x$, $x_\chi$, and $z$ for $\sigma = 0.75$ and $\sigma = 0.85$ are summarized in Table II for both the XY and Heisenberg models.

The data collapse is shown in Figs. S2(a) and S2(b). We determine $z$ by noting that when $N^{1/x} h_r$ is large, $\chi_{\rm SG}$ becomes independent of $N$. Remarkably, the value of $z$, which fits well at large $N^{1/x} h_r$, also provides a good collapse for small $N^{1/x} h_r$. The distinct behaviors for $\sigma = 0.75$

and $\sigma = 0.85$ at the largest system sizes are evident in Fig. S3, displayed on a log-log plot to show behavior across all values of $h_r N^{1/x}$ studied. Corrections to the Imry-Ma scaling form are very noticeable, in particular in Fig. S2(a) for small $N^{1/x} h_r$, similar to the behavior in the $\xi_{\mathrm{SG}}$ plot, Fig. 2(a), which are both for $\sigma = 0.75$, and due to RSB effects.


\fi

%==========================================================================

\end{document}